\documentclass[twocolumn,showpacs,amsmath,amstex,amssymb,mathfonts]{revtex4}
\usepackage{graphicx,epstopdf,verbatim,amsmath, color}

\begin{document}
\title{Non-local response in the disordered Majorana chain}
\author{Sonika Johri$^1$ and Rahul Nandkishore$^{2,3}$}
\affiliation{$^1$Quantum Computing Group, Intel Labs, Intel Corporation, Hillsboro, OR 97214\\
		$^2$Department of Physics, University of Colorado, Boulder, CO 80309, USA\\
		$^3$ Center for Theory of Quantum Matter, University of Colorado, Boulder, CO 80309, USA}
	
\begin{abstract}
The prospects for realizing a topological quantum computer have brightened since the apparent detection of Majorana fermions at the ends of semiconducting nanowires. These Majorana zero-modes persist in the presence of the strong disorder that may be present in such systems, protected by the mobility gap of localized systems. Recent proposals to manipulate quantum information in a Majorana chain involve adiabatically adjusting gate voltages in one-dimensional nanowire networks. However, as we show, in the adiabatic limit, a disordered system with multiple zero-mode Majorana fermions will undergo a non-local response to a local perturbation via a multi-level Landau-Zener transition. This will disperse the quantum information stored in the target zero-modes amongst other zero-modes that are present in the system, complicating the realization of controlled gates. There is however a solution that may work in practice: we show that there is an optimum window of intermediate time-scales during which the quantum information is approximately preserved under manipulation. %The non-local response persists even when one accounts for the fact that the number of zero modes is conserved only modulo two.
\end{abstract}
	
\date{\today}
\maketitle

Majorana fermions satisfy non-Abelian braiding statistics which means that they can act as the building blocks of a topological quantum computer\cite{nayak}. In condensed matter settings, Majorana fermions (denoted in this paper by $\gamma_j$) occur in pairs, arising from the decomposition of Dirac fermion operators, $a=\gamma_1+i\gamma_2$. Kitaev proposed a theoretical model in which zero-energy Majorana modes occur at the ends of a one-dimensional quantum chain with superconducting coupling\cite{kitaev}. Further, one can also show that Majorana zero-modes would be realized at boundaries between topologically distinct phases in such chains. Recent experiments claim to have realized this model in semiconducting nanowires and detected the existence of Majorana zero-modes at the edges\cite{mourik}-\cite{nadj-perge}. The next step towards realizing topological quantum computing is to implement braiding operations and confirm the non-Abelian statistics of these quasiparticles. To this end, a protocol which uses a network of these nanowires and implements braiding by simply varying gate voltages has been proposed\cite{alicea}. In addition to the zero-modes at the edges, the protocol also utilizes Majorana zero-modes in the bulk which exist at domain walls between trivial and topological zones. These domain walls are also created by varying a series of gate voltages over the spatial extent of the system. Central to this technique is the assumption that keeping all operations adiabatic with respect to the superconducting gap to excited states preserves the fidelity of the quantum information.

In realistic systems, disorder effects need to be taken into account. Strong enough disorder gives rise to localization accompanied by a gapless spectrum. In this scenario, it is the mobility gap of localized systems which serves to protect Majorana zero-modes, analogous to the role played by the superconducting gap in the clean system \cite{LPQO}. The disorder exponentially localizes zero-modes with localization length controlled by the disorder strength. The presence of disorder also gives rise to the possibility of additional zero modes being present at random in the bulk of the system. However, it has recently been pointed out that adiabatic manipulations in gapless localized systems are potentially problematic, because of the problem of non-local response, whereby perturbations on a timescale $\tau$ induce a response on a length scale $\xi \ln(\tau)$, where $\xi$ is the localization length \cite{khemani}. While this result greatly constrained the possibilities for quantum control and manipulation of quantum information in e.g. quantum Hall systems, it is not clear whether it applies to the Majorana chain, for two reasons. Firstly, the calculations of \cite{khemani} were for a model with a conserved particle number, whereas in the Majorana chain particle number is conserved only modulo two. Additionally, the Majorana chain is a symmetry protected topological (SPT) phase, and such phases have also been argued to have special properties under dynamical manipulations \cite{vishwanath}. Thus, the question of whether a safe adiabatic limit exists for dynamical manipulations in the gapless localized phase of the Majorana chain remains open.

In this paper we will show by means of simulations based on numerical exact diagonalization that the adiabatic limit is not suitable for manipulating zero modes in gapless Majorana chains protected by localization. In general, a time-dependent local potential applied to this system will lead to a multi-level Landau-Zener transition \cite{multi_LZ_1, multi_LZ_2, multi_LZ_3} which will decrease the fidelity of the operation. %The multi-level nature of the problem arises because Majoranas occur in pairs with energies $\pm E$.
%A time-dependent local potential leads to a non-local response which induces oscillations in the fidelity of the qubit. Further, when there are several pairs of zero-modes in the bulk of the system, the recovery time for the fidelity, in other words the time period of the oscillations, may reach infinity. 
This work closes the twin loopholes discussed above, and establishes that the problem of non-local response to local manipulations \cite{khemani} is a general one, which applies even to SPT systems without a conserved $U(1)$ charge such as the Majorana chain. It follows that there is no safe adiabatic limit for dynamic manipulations in the Majorana chain. However, depending on the disorder realization, there may be an intermediate range of timescales over which dynamic manipulations with acceptable (but not asymptotically good) fidelity may be performed.

The system is described by a time dependent Hamiltonian of the form $H = H_0 + V(t)$, where $H_0$ is the time independent Hamiltonian of a gapless localized system and $V (t)$ is a local (in real space) time dependent perturbation. It is sufficient for our purposes to consider a non-interacting localized system in one dimension with open boundary conditions. The time independent piece of the Hamiltonian is
\begin{align}\label{eq:H0}
H_0=i\sum_{j=1}^{N/2-1}J_j\gamma_{2j}\gamma_{2j+1}+i\sum_{j=1}^{N/2}h_j\gamma_{2j-1}\gamma_{2j},
\end{align}
where $\gamma_j$ are Majorana operators which anti-commute and square to 1, and the $J_j$ and $h_j$ are random coupling constants drawn from box distributions with mean $\bar{J}$ and $\bar{h}$ and width $\Delta_J$ and $\Delta_h$ respectively. We simplify by taking $\Delta_h = 0$ and set all $h_j = \bar{h}=2$. This is sufficient for our purposes. The topological phase is $\bar{J} > ̄\bar{h}$, so we take $J_j$ to be drawn from a box distribution $[0, 2\bar{J}]$ with $\bar{J}=20$. $\bar{J}>>\bar{h}$ ensures that at least two zero-energy Majorana modes exist at the ends of the chain. The length $N$ of the chain is constrained to be an even number because Majoranas are created in pairs from the physical superconducting system.

The matrix $H_0$ can be written in the form
\begin{align}
H_0=\frac{i}{2}\sum_{j=1}^{N/2}\epsilon_j\tilde{\gamma}_{2j-1}\tilde{\gamma}_{2j}= \sum_{j=1}^{N/2}\epsilon_j\bigg(c^{\dagger}_jc_j-\frac{1}{2}\bigg),
\end{align}
where the new Majorana operators $\tilde{\gamma}_j=\sum_i W_{jk}\gamma_k$. New Dirac fermion operators $c_j^{\dagger}=(\tilde{\gamma}_{2j-1}-i\tilde{\gamma}_{2j})/2$ and $c_j=(\tilde{\gamma}_{2j-1}+i\tilde{\gamma}_{2j})/2$ can also be defined which diagonalize $H_0$. Numerically, we use exact diagonalization of the random matrix described by $H_0$ in Eq. \ref{eq:H0} to obtain the values $\epsilon_j$ \cite{numerical}. When one or more $\epsilon_j$ vanish, we obtain Majorana zero-modes. For $\bar{J}>>\bar{h}$, there is one pair of zero-energy modes localized at at each end of the chain generated by operators $\gamma_1 + \frac{h_1}{J_1}\gamma_3+\frac{[h_1h_2]}{J_1J_2}\gamma_5+...$ and $\gamma_N + \frac{h_N}{J_N}\gamma_{N-2}+...$. In addition, there will also be Majorana zero modes at any place where there is a break in the topological phase (i.e. where locally $J < h$). For a finite-size system or a finite density of quasi-zero modes, the zero-modes will not have exactly zero energy but have energy values which are exponentially small in the spacing between quasi zero modes.

We want to simulate the protocol proposed in Ref. which utilizes several zero-modes localized at different points in the system. To do this, we deliberately create extra pairs of zero-modes by setting $h$ to be a series of step functions:
\begin{equation}\label{h_j}
h_j=
\begin{cases}
\bar{h}, & \text{if } 0<j<j_1, j_2<j<j_3,\ldots, j_{n-2}<j<N\\
h_{T}, & \text{else}.
\end{cases}
\end{equation}
where $h_{T}=\max(J_j)$. Now, with high probability there will be zero-modes localized at $0$, $j_1$, $j_2\ldots$, $N$ where there is a domain wall between trivial and topological phases. Since the expected zero-modes are separated by a finite distance, the system is only quasi-degenerate with $n$ modes that have energies close to $0$. Numerically, we enforce that the disorder configuration we are studying has the intended number of zero modes by checking that exact diagonalization produces $n$ energies with magnitude less than $0.1$ times the average energy spacing $2\bar{J}/N$ between eigenvalues of the disordered system. We denote the zero-energy modes by $\phi_k$ where $k=1,2,\ldots,n$.

\begin{figure}
	\includegraphics[width=\columnwidth]{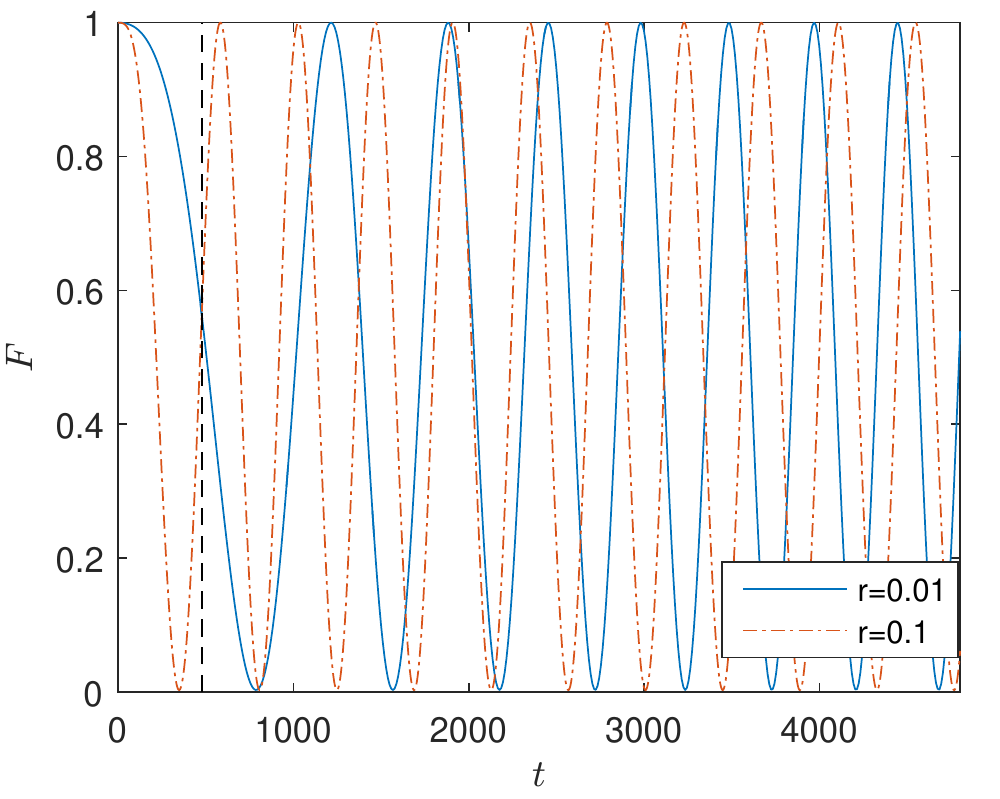}
	\caption{Fidelity $F$ versus time $t$ for a disordered chain of $N=100$ Majoranas under a linear ramp of $h_1$ from $\bar{h}$ to $h_{\max}$. The chain contains two zero-modes in the bulk created by engineering $h_j$ according to Eq. \ref{h_j}. The dashed black vertical line indicates the time when the ramp is turned off.}
	\label{fig:fidelity_4}
\end{figure}

Now we want to locally manipulate (move around) the Majorana mode at one end of the chain. We do this in the manner of Ref. 2 by applying a local field $V (t) = h_1(t)=rt$. We ramp up $h_1$ linearly from its original value $\bar{h}$ to a large final value such as $h_{\max} = 2.5\bar{J}$ over a timescale $\tau=h_{\max}/r$. We use a fourth-order Runge-Kutta method to perform numerical time evolution. The expected outcome of applying $V(t)$ adiabatically is that the zero-mode localized at $j=1$ moves inwards and is now localized at $\sim j=3$.

We note that the Hamiltonian (2) is quadratic in fermion operators so it is sufficient to consider the time evolution of a single particle wavefunction $\Psi$. At time $t$ the wavefunction $\Psi$ has amplitude $\Psi_j$ at site $j$ of the lattice. We start with $\Psi_1=1$ and all other $\Psi_j=0$ at time $t=0$. We note that physically one would start with a wavefunction $\Psi=\phi_{k_1}+i\phi_{k_2}$, $k_1\neq k_2$, but that would not change the results of this work. 

We define fidelity by how closely the outcome of the manipulation adheres to the expected result, namely that the weight of the wavefunction should move from $j=1$ to $j=3$ and the wavefunction should stay in the zero-energy subspace. Quantitatively, fidelity
\begin{align}\label{eq:fidelity}
F=\sum_{k}\sum_{j=1}^{j_{\text{cut}}} B_k |\phi_{k,j}|^2
\end{align}
where, $B_{k}=|\langle\Psi|\phi_k\rangle|^2$ is the overlap of the wavefunction with the $k$th zero-mode and $\phi_{k,j}$ is the amplitude of the $\phi_k$ on the $j$th site of the lattice. We choose $j_\text{cut}=7$ in order to provide reasonable tolerance for including the exponential tail of the localized states. We check that $F>0.99$ at $t=0$. This ensures that the disorder configurations chosen satisfy the condition of having zero-energy modes at the ends of the chain.

We also define the expectation value of spatial position of the wavefunction 
\begin{align}
 L=\sum_{j=1}^{N} j |\Psi_j|^2
\end{align}
$L$ quantifies the non-local response to a local potential applied at the left end of the chain.

\begin{figure}
	\vspace{-30 pt}
	\includegraphics[width=\columnwidth]{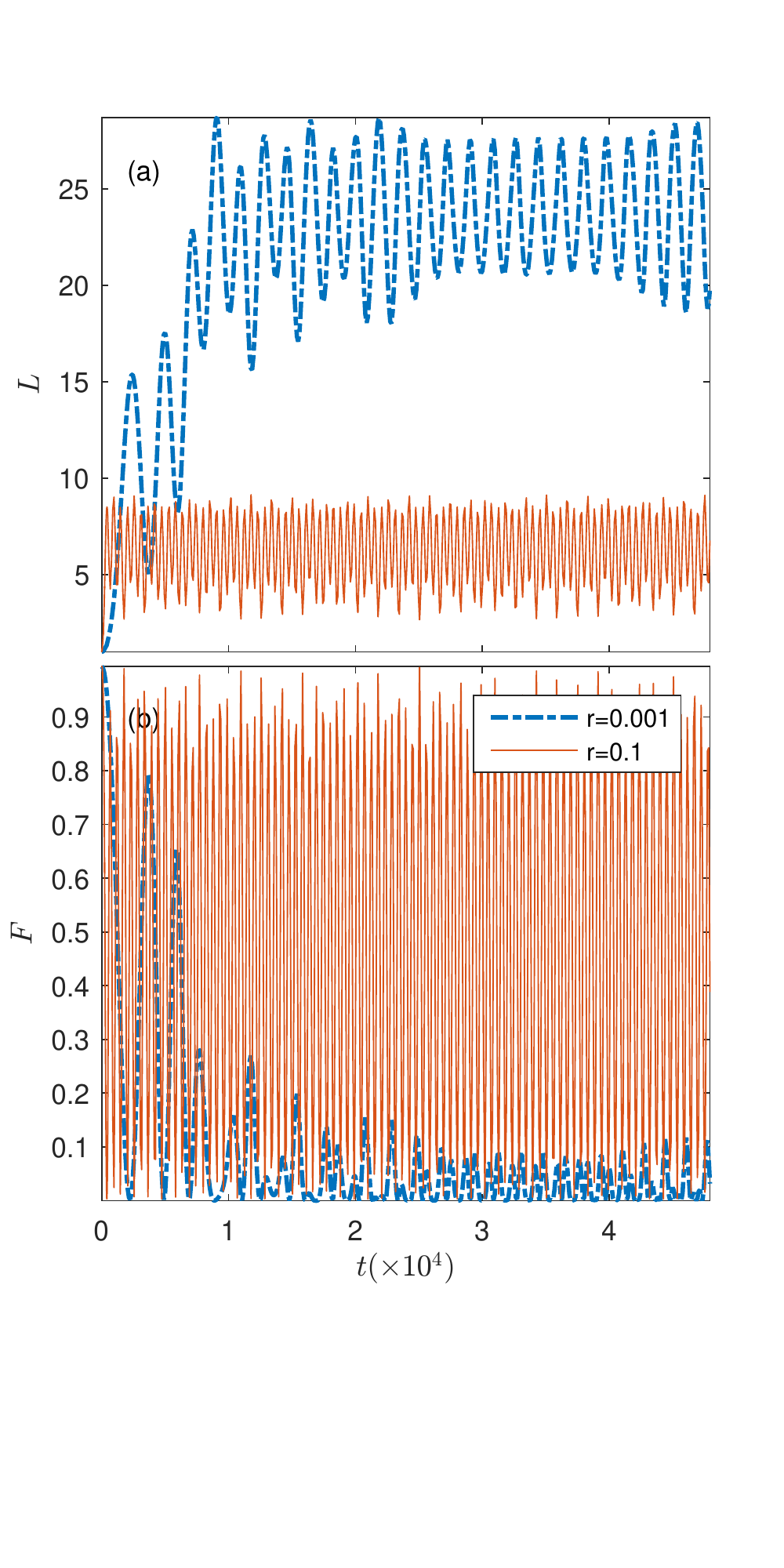}
	\vspace{-100 pt}
	\caption{(a) Expectation value of position $L$, and (b) fidelity $F$, versus time $t$ for a disordered chain of $N=100$ Majoranas for two different values of the ramp rate $r$. The chain contains six zero-modes in the bulk created by engineering $h_j$ according to Eq. \ref{h_j}. \label{fig:fidelity_loc_several_vary_dh}
}	
\end{figure}

Let us analyze what happens when $h_1$ is increased. Fig. \ref{fig:fidelity_4} shows the evolution of fidelity $F$ as $h$ is increased from $\bar{h}$ to $h_{\max}$ for a chain of $N=100$ Majoranas with 4 zero-modes localized at sites $j=0, 10, 90, 100$. Instead of remaining close to $1$, the fidelity oscillates between $0$ and $1$. Reducing the ramp ramp rate $r$ by an order of magnitude preserves the oscillations. The reason for this becomes clear upon the realization that $V(t)$ couples the Majorana at the left end with the zero-mode at $j=10$. The frequency of the oscillations will depend on the distance between the zero-modes and the ramp rate $r$. 
%Thus the fidelity drops because the local perturbation has generated a non-local response. 
While a zero-mode at the edge is maintained, the wavefunction itself evolves in a quasi-degenerate subspace. In the infinitely slow limit, any quantum information stored in the system should stay confined within the quasi-degenerate zero-energy subspace. However, this does not mean that the fidelity as defined above is preserved since the information can disperse amongst other zero-energy modes located far away. %This means that the adiabatic limit for manipulations will not protect quantum information stored in the system.

\begin{figure}\label{fig:fidelity_loc_several_ramp_hold}
	\includegraphics[width=\columnwidth]{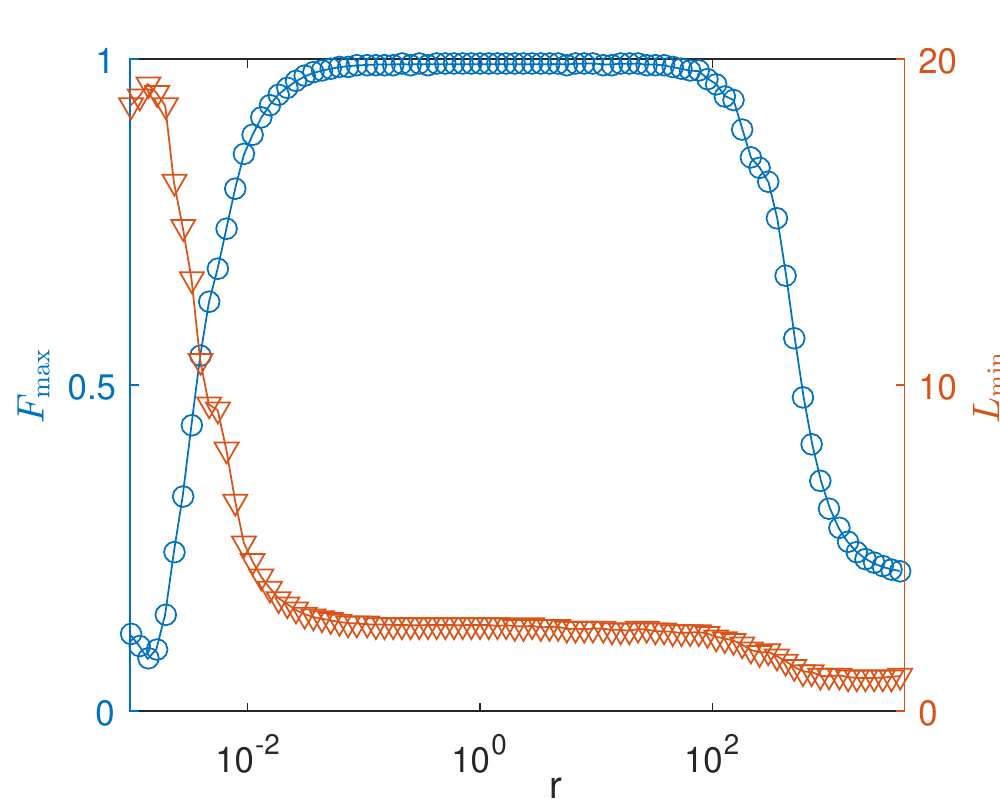}
	\caption{The maximum fidelity $F$ and minimum $L$ after the ramp is over versus ramp rate $r$ for the disordered chain of $N=100$ Majoranas in Fig. \ref{fig:fidelity_loc_several_vary_dh} with six zero modes in the bulk.}
\end{figure}

In this situation, it is clear that though fidelity falls to 0, it recovers after some time. Is this also the case when several zero-modes are present in the bulk? Figure 2 shows $F$ and $L$ for different ramp rates $r$ for a Majorana chain of length $N=100$ with 8 zero-modes localized at $j=0, 10, 20, 30, 70, 80, 90 ,100$. At the larger value of $r$, the time evolution exhibits the phenomenon of beating between oscillations of different frequencies. For $r$ two orders of magnitude smaller, in addition to beating, the fidelity drastically reduces during the ramp and correspondingly the non-local response $L$ increases indicating that the quantum information stored in the edge Majorana has dispersed over the zero-modes in the bulk.

Does the fidelity in this case ever recover? Fig. 3 shows the maximum value of $F$ and minimum value of $L$ obtained during the evolution of the system for a time interval of length $\Delta t=5000$ after the end of the ramp, as a function of $r$ for the disorder configuration used in Fig. 2. There are three discernible regions in the figure. The center consists of the optimal range of ramp rates during which maximum fidelity is close to 1. At large $r$, the fidelity drops because any quantum information stored in the zero-energy subspace will disperse into the excited states. This is the well-understood diabatic error, of which a comprehensive recent discussion can be found in \cite{Knappetal}. The lower bound to this region will be determined by the mobility gap. The recoverable fidelity also decreases as $r$ is decreased at low values of $r$. The non-locality of the response measured by $L$ in this region is high. This permanent reduction in the fidelity at small $r$ occurs due to a Landau-Zener transition within the quasi-zero energy subspace (see Appendix A for a plot of the energy levels). The adiabaticity condition
\begin{align}\label{eq:adiabaticity}
\frac{\langle\phi_a|\frac{\partial V}{\partial t}|\phi_b\rangle}{(E_a-E_b)^2}<<1.
\end{align}
determines whether the quantum information will remain localized in the edge mode. Here, the energy difference between zero-modes $(E_a-E_b)$ will be exponentially small in the distance between them. If $\tau$ is of the order of $(E_a-E_b)$, the system can remain adiabatic. This gives the same length - time scaling $R \sim \ln \tau$ that was identified in \cite{khemani}. 

\begin{figure}\label{fig:fuse}
	\includegraphics[width=\columnwidth]{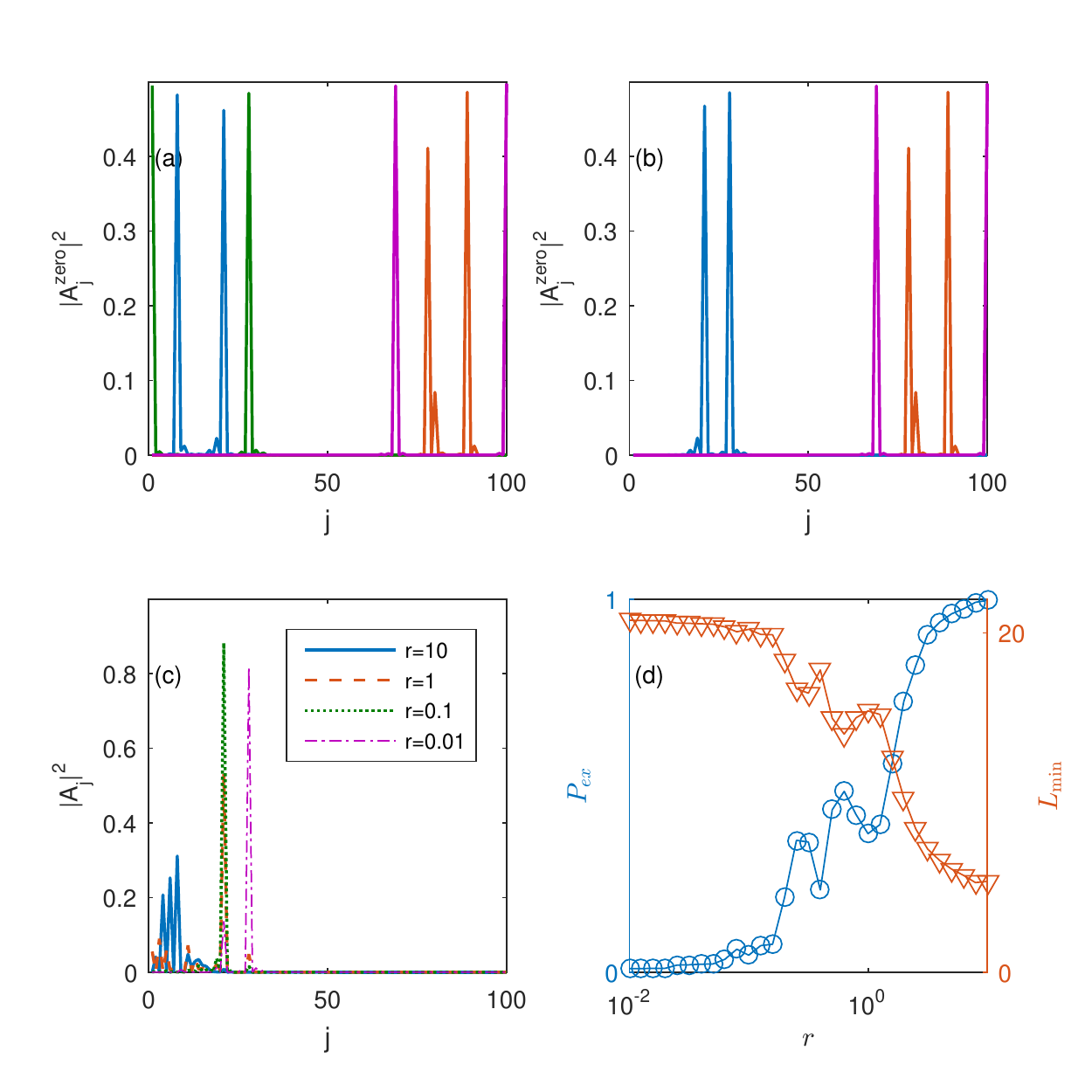}
	%\vspace{-100 pt}
	\caption{Amplitudes of quasi-zero energy modes in the lattice (a) before and (b) after the time-dependent perturbation. (c) The final wavefunction for different values of the ramp rate $r$. (d) Probability $P_{\text{ex}}$ of the wavefunction escaping the zero-energy subspace and minimum value of $L$ after the end of the ramp as a function of $r$.}
\end{figure}

To get an estimate for the probability of successful manipulation, we ran the above protocol for different disorder configurations. We defined a successful manipulation if the system manages to reach a value of $L\geq 2.8$ and $F>=0.99\times F(t=0)$ while $h_1$ is changed from $\bar{h}$ to $h_{\max}=2.5\bar{J}$ at ramp rate $r=0.01$. Out of 100 disorder configurations, no zero-modes in the bulk recorded success probability of 97\% of the time, for 2 zero-modes (located at $j=10, 90$) 82\% of the time, and for 4 zero-modes (located at $j=10, 20,80,90$) 35\% of the time. Thus the success probability for the protocol in \cite{alicea} falls rapidly as the number of zero-modes in the bulk is increased if the time-scale is not chosen properly. We have also checked whether a cycle involving ramping up, holding and then ramping down $h_1$ similar to the one used in \cite{vishwanath} would restore the fidelity but it does not.

Next we look at what happens when the number of zero-modes in the ground state changes, i.e. U(1) charge conservation is violated. We fuse two zero-modes together by successively increasing $h_1$, then $h_2$, all the way to $h_5$ from their initial value of $\bar{h}$ to $h_{\max}$. Fig. 4(a) shows the 8 initial zero-energy modes before the procedure and Fig. 4(b) shows the 6 final zero-energy modes. Note that the different colours in the figure indicate that the eigenstates output from the numerical exact diagonalization are superpositions of zero-modes. We start with the same initial wavefunction as before. Fig. 4(c) shows the final wavefunction for different ramp rates. In Fig. 4(d), the probability of the quantum information escaping into the excited states $P_{\text{ex}}$ and the minimum value of $L$ obtained during a time evolution for an interval $\Delta t$ after the end of the ramp are shown as a function of $r$. Their behavior indicates that Landau-Zener physics is again in play here. %, and that in the adiabatic limit $r \rightarrow 0$ the system exhibits a non-local response. 
For large $r$, the wavefunction does not migrate and moves into the excited states indicating a diabatic process. For slow enough ramp rates, the process remains adiabatic and diffuses to the nearest zero-energy mode. As $r$ is further decreased, the amount of the wavefunction that stays in the zero-energy subspace increases as can be deduced from the adiabaticity condition in Eq. \ref{eq:adiabaticity}, but the response becomes increasingly non-local i.e. information disperses through the system because of Landau Zener transitions. We note that $P_{\text{ex}}$ and $L_{\min}$ are oscillatory functions of $r$. The beginning of an oscillation can also be observed near the very left edge of the curves in Fig. 3. This arises because the Majorana nature of the zero-modes leads to a multi-level Landau-Zener transition \cite{multi_LZ_1, multi_LZ_2, multi_LZ_3} instead of the typical 2 level transition. See Appendix A for a discussion. For completeness, we have also tested the protocol in \cite{alicea} in the presence of finite-energy excitations in the bulk (See Appendix B). For the non-interacting model considered here there is no difference in the behavior.

We conclude that the nonlocal response of localized systems to local perturbations \cite{khemani} generalizes also to settings without particle number conservation and to SPT systems. Important observations are that avoided crossings can happen in the quasi zero-energy subspace, that they happen in the absence of particle number conservation, that the nature of the Majoranas may sometimes cause the crossings to be multi-level (leading to oscillatory behavior in the fidelity as a function of ramp rate). 

 As a result the adiabatic protocol advanced in \cite{alicea} is unsuitable for manipulating Majorana chains with several zero-modes. Zero-modes in the bulk may occur by chance in disordered systems or may be deliberately created in order to increase the quantity of quantum information that can be stored in the system. However, doing so may increase the probability of failure. It is also clear that there do exist time scales for which the quantum information can be safely manipulated. However, there is no safe asymptotic limit for braiding, and the window of timescales that gives acceptable fidelities depends on the particular disorder realization. 

\appendix
\renewcommand\thefigure{\thesection.\arabic{figure}} 
\section{Landau Zener transitions in the Majorana chain}
\setcounter{figure}{0}  

Fig. A.1 shows how the positive energy levels for the disorder configuration in Fig. 3 change under the ramp on $h_1$. There is clearly an avoided level crossing that occurs around $h_1=6$. Each positive energy level has a negative counterpart and the avoided level crossing thus occurs in the negative energies as well.

\begin{figure}\label{fig:energies}
	\includegraphics[width=\columnwidth]{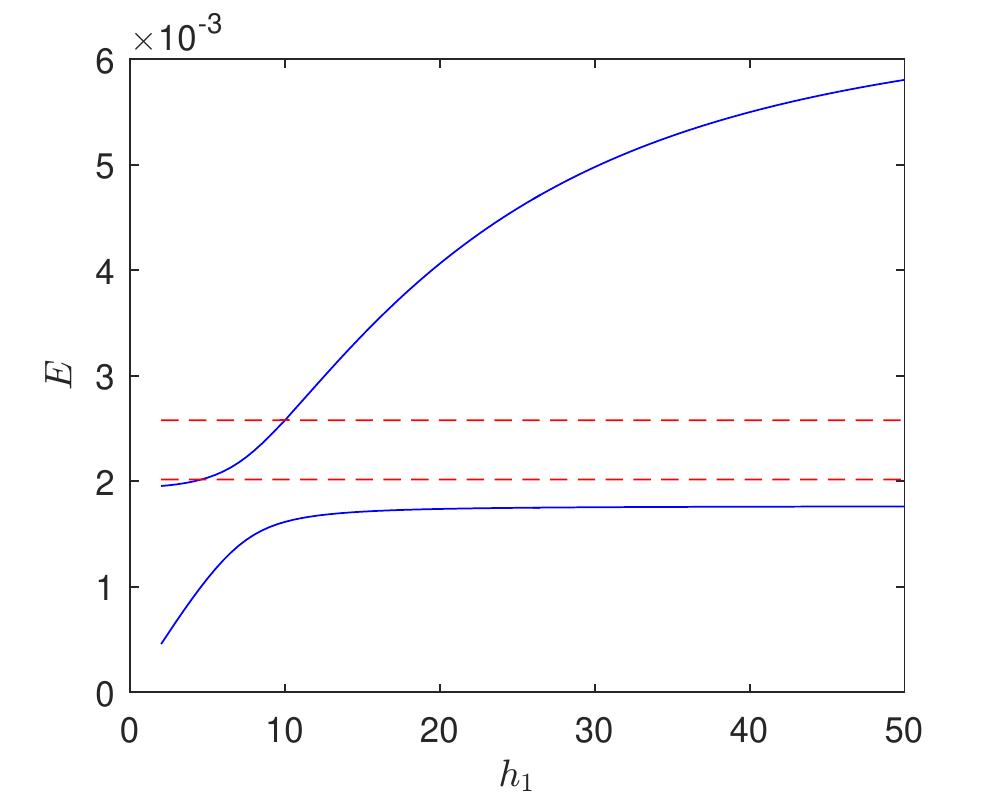}
	\caption{An avoided crossing for the energy levels for the disorder configuration in Fig. 3. The zero-modes with positive energies are displayed. The two energies that are unchanged (red dashed lines)are located at the other end of the chain and the two time-dependent energies (blue solid curves) are located close to the perturbing field. In spite of the energies being confined to the quasi-zero-energy subspace by the topological nature of the phase (Note the $10^{-3}$ multiplier on the y-axis.), there is still an avoided crossing.}
\end{figure}

Next we discuss the multi-level nature of the transition which leads to the oscillations in Fig. 4. We illustrate this by way of considering 3 zero-mode Majoranas $\gamma_1$, $\gamma_2$ and $\gamma_3$ arranged from left to right near the left end of the chain with a fourth zero mode far away at the right end of the chain. Their overlap with each other is exponentially small in the distance between them but it is not zero. Note that one of the eigenvalues is forced to be almost exactly zero in this case because of the symmetry between positive and negative energies. The eigenstates will be some combination of $\gamma_1$, $\gamma_2$ and $\gamma_3$.

As the boundary field $h_1$ is ramped up, $\gamma_1$ approaches $\gamma_2$ and the eigenenergies split into $\pm E$ with accompanying eigenstates which now have equal amplitude on both $\gamma_1$ and $\gamma_2$. $|\langle\Psi_+|V|\gamma_3\rangle|$ and $|\langle\Psi_-|V|\gamma_3\rangle|$ should then be equal (where $V$ is the ramp, and $\Psi_+$ and $\Psi_-$ correspond to $\pm E$ eigenstates). Therefore, we have the situation where $\gamma_1$ and $\gamma_2$ are strongly coupled and $\gamma_3$ is weakly coupled to both of them. Therefore this is a 3 level problem leading to the oscillatory behavior of fidelity as a function of ramp rate shown in Fig.4. 

\section{Finite Energy Excitations}
\setcounter{figure}{0}  

We checked the efficacy of the protocol in \cite{alicea} when finite energy excitations are present in the system. We start with a system that in addition to 4 pairs of zero-modes also has finite-energy excitations present at time $t=0$. The ramp on $h_1$ is then turned on till $h_1=h_{\max}$. The fidelity under two different ramp rates is shown in Fig. B.1. Here we plot the normalized fidelity because the fidelity as defined in Eq. \ref{eq:fidelity}  at time $t=0$ is significantly less than 1 because part of the wavefunction at $t=0$ lies in the excited energies. We test the chain with two excitations present close to the left end of the chain. In both cases, the fidelity is not reduced from the case with no excitations. This may however be an artifact of the non-interacting model being simulated here. 

\begin{figure}\label{fig:en_finite}
	\includegraphics[width=\columnwidth]{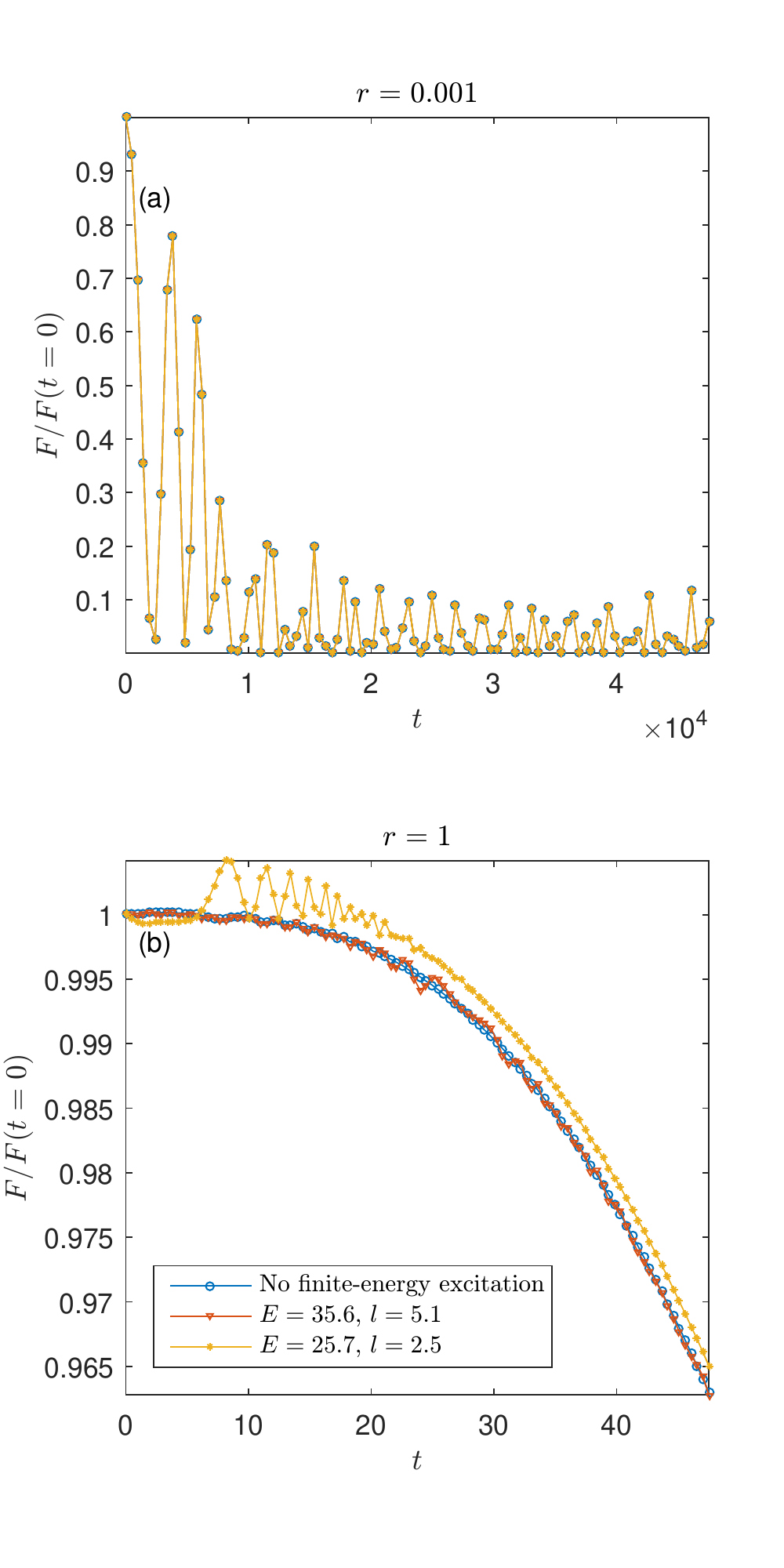}
	\caption{Normalized fidelity versus time $t$ for a disordered chain of $N=100$ Majoranas under a linear ramp of $h_1$ from $\bar{h}$ to $h_{\max}$ when finite energy excitations are present at $t=0$ at (a) $r=0.001$ and (b) $r=1$. The chain contains four pairs of zero-modes created by engineering $h_j$ according to Eq. \ref{h_j}. It also contains a finite energy excitation with energy $E$ and expectation value of position $l$ listed in the legend.}
\end{figure}

\end{document}